\documentclass[preprint,12pt,showpacs]{revtex4}

\usepackage[brazil, english]{babel}
\usepackage[utf8]{inputenc}
\usepackage{graphicx}
\usepackage{epsfig}
\usepackage{amssymb}
\usepackage{amsmath}
\usepackage{slashed}

\graphicspath{%
    {converted_graphics/}
    {/}
}
\begin{document}

\title{Dynamical corrections to the anomalous holographic softwall model: the pomeron and the odderon}
\author{Eduardo Folco Capossoli$^{1,2,}$}
\email[Eletronic address:]{educapossoli@if.ufrj.br}
\author{Danning Li$^{3,}$}
\email[Eletronic address:]{lidn@itp.ac.cn}
\author{Henrique Boschi-Filho$^{1,}$}
\email[Eletronic address: ]{boschi@if.ufrj.br}  
\affiliation{$^1$Instituto de F\'{\i}sica, Universidade Federal do Rio de Janeiro, 21.941-972 - Rio de Janeiro-RJ - Brazil \\
 $^2$Departamento de F\'{\i}sica, Col\'egio Pedro II, 20.921-903 - Rio de Janeiro-RJ - Brazil \\ 
 $^3$ Institute of Theoretical Physics, Chinese Academy of Science (ITP, CAS), 100190 - Beijing - China}

\begin{abstract}
In this work we use the holographic softwall  AdS/QCD model with anomalous dimension contributions coming from two different QCD beta functions to calculate the masses of higher spin glueball states for both even and odd spins and its respective Regge trajectories, related to the pomeron and the odderon, respectively. We further investigate this model taking into account dynamical corrections due to a dilaton potential consistent with Einstein equations in 5 dimensions. The results found in this work for the Regge trajectories within the anomalous softwall model with dynamical corrections are consistent with those presented in the literature.
\end{abstract}


\maketitle


\section{Introduction}

Quantum Chromodynamics (QCD) is a non-Abelian Yang-Mills gauge theory that correctly describes the strong interactions. Its well known Lagrangian is given by:
\begin{equation}\label{lqcd}
{\cal L}_{{\rm QCD}} = {\bar{\psi}} \left( {\slashed D} - m\right) \psi - \frac{1}{4} G^a_{\mu \nu} G^{\mu \nu}_a \;,
\end{equation}
\noindent where $\psi (x)$ represents the quark fields in the fundamental representation of the $SU(3)$ gauge group and $G^a_{\mu \nu}$ is the gluon field strength tensor. This tensor can be decomposed in a such way, as
\begin{equation}\label{gst}
G^a_{\mu \nu} = \partial_\mu \mathcal{A}^a_\nu - \partial_\nu \mathcal{A}^a_\mu + g_{YM} f^{abc} \mathcal{A}^b_\mu \mathcal{A}^c_\nu \,,
\end{equation}
\noindent where $\mathcal{A}^a_\nu$  are the gluon fields, with $a = 1, \cdots, 8$, $f^{abc}$ are the structure constants of $SU(3)$ group and $g_{YM}$ is the coupling constant of Yang-Mills (strong) interactions. 

Gluons do not carry electric charges, but they have color charge. Due to this fact, they coupled to each other, as implied by Eqs. (\ref{lqcd}) and (\ref{gst}). The bound states of gluons predicted by QCD, but not detect so far, are called glueballs. 
Glueballs states are characterised by $J^{PC}$, where $J$ is the total angular momentum, and $P$ and $C$ are the $P-$parity (spatial inversion)  and  the $C-$parity (charge conjugation) eigenvalues, respectively. 

QCD has many interesting characteristics and among them asymptotic freedom and confinement. The first one means that in the ultraviolet limit, i.e, for high energies or short distances, quarks and gluons practically do not interact with each other. In this case, one has a weak coupling regime and QCD can be treated perturbatively. On other hand, confinement means that in the infrared limit, i.e, for low energies or large distances, quarks and gluons are bounded to each other strongly, in a strong coupling regime,  and therefore, inaccessible to the perturbative approach. Calculations involving bound states as the glueball masses, and consequently, its related Regge trajectories, are  features of the non-perturbative regime. 
Regge trajectories are well known approximate linear relations between total angular momenta $(J)$ and the square of the masses $(m)$, such that:
\begin{equation}
J(m^2) \approx \alpha_0 + \alpha' m^2 \;,
\end{equation}
\noindent with $\alpha_0$ and $\alpha'$ constants.

In this paper we will calculate the masses of the even and odd spin glueball states and from this masses one gets the Regge trajectories for the pomeron and odderon, respectively. As mentioned before those glueball states were not detected experimentally so far. Due to this, the glueball state masses come from some particular models. In the case of the pomeron, related to the even spin glueball state masses, there are some models to describe it, such as, the soft pomeron \cite{Landshoff:2001pp}, given by:
\begin{equation}\label{pomeron}
J(m^2) \approx 1.08 + 0.25 \, m^2 \;,
\end{equation}
where the masses throughout this work are expressed in GeV. There are also other models like 
the triple pole pomeron \cite{Cudell:2001ii} , the hard pomeron \cite{Levin:1998pk}, the BFKL pomeron \cite{Levin:1998pk} and etc. All of these models provide trajectories pretty close to eq. \eqref{pomeron}, so we are going to 
compare our results obtained for the Regge trajectory of the pomeron with this one.  

In the case of odderon, related to odd spin glueball states, there are also many models to describe it, such as, isotropic lattice \cite{Meyer:2004jc}, anisotropic lattice \cite{Chen:2005mg}, relativistic many body \cite{LlanesEstrada:2005jf}, given by:
\begin{equation}\label{r_odderon}
J(m^2) \approx -0.88 + 0.23 \, m^2 \;,
\end{equation}
the non-relativistic constituent \cite{LlanesEstrada:2005jf} given by: 
\begin{equation}\label{nr_odderon}
J(m^2) \approx 0.25 + 0.18 \, m^2 \;,
\end{equation}
and etc. We are going to compare our results obtained for the Regge trajectory of the odderon, with these two trajectories. 

The AdS/CFT or Anti de Sitter/Conformal Field Theory correspondence \cite{Maldacena:1997re, Gubser:1998bc, Witten:1998qj, Witten:1998zw, Aharony:1999ti} arises as a powerful tool to tackle non-perturbative Yang-Mills theories. The AdS/CFT correspondence relates a conformal Yang-Mills theory with the symmetry group $SU(N)$ for very large $N$ and extended supersymmetry $({\cal N} = 4)$  with a $IIB$ superstring theory in a curved space, known as anti de Sitter space, or $AdS_5 \times S^5$. At low energies string theory is represented by an effective supergravity theory, due to this reason the AdS/CFT correspondence is also known as gauge/gravity duality.

After a suitable breaking of the conformal symmetry one can build phenomenological models that describe (large $N$) QCD approximately. These models are known as AdS/QCD models.

In order to deal with conformal symmetry breaking the works \cite{Polchinski:2001tt, Polchinski:2002jw, BoschiFilho:2002ta, BoschiFilho:2002vd} have done some important progress with this issue. In these two last works, emerges the idea of the hardwall model. This idea means that a hard cutoff is introduced at a certain value $z_{max}$ of the holographic coordinate, $z$, and is considered a slice of $AdS_5$ space in the region $0 \leq z  \leq z_{max}$, with some appropriate boundary conditions. 

Another holographic approach to break the conformal invariance in the boundary field theory, and make it an effective theory for large $N$ QCD, is called the softwall model. This model  introduces in the action a decreasing exponential factor of the dilatonic field that represents a soft IR cutoff. The softwall model (SW) was proposed  in \cite{Karch:2006pv} to study vector mesons, and subsequently extended to  glueballs \cite{Colangelo:2007pt} and to other mesons and baryons \cite{Forkel:2007tz}. It was shown in the ref. \cite{Capossoli:2015ywa} that the softwall model does not give the expected masses for the scalar glueball states (and its radial excitations) and higher spin glueball states (with even and odd spins). 

In ref. \cite{Gursoy:2007cb}, it was introduced the idea of using QCD beta functions to get an interesting UV behaviour for the softwall model modified by convenient superpotentials for the dilaton field. 
Then, in reference \cite{BoschiFilho:2012xr} it was proposed a simpler modification of the SW model, taking into account  anomalous dimensions, also related to QCD beta functions to obtain the scalar glueball spectra and its (spin 0) radial excitations. The resulting masses found for some choices of beta functions are in agreement with those presented in the literature. 

In this work, our main objective is to extend the previous studies done in ref.\cite{BoschiFilho:2012xr} to investigate higher, even and odd, spin glueball states, and obtain the Regge trajectories related to the pomeron and to the odderon, taking into account the anomalous dimensions from some QCD beta functions with  dynamical corrections, i.e, considering that the dilatonic field became dynamical satisfying the Einstein equations in five dimensions. Actually, we first consider the anomalous contributions to the original softwall model without dynamical corrections for higher even and odd spins, but the results are not good compared with the pomeron and odderon Regge trajectories. Then, we move to consider the dynamical case where the results obtained are good. 

This work is organised as follows: In the section \ref{ASW} we provide a quick review of the original softwall model (SW) and the modifications taking into account the anomalous dimensions from QCD beta functions that we call the anomalous softwall model (ASW) (that is the softwall with anomalous contributions as proposed in ref.\cite{BoschiFilho:2012xr}). Then, we introduce even and odd higher spin state operators and calculate, from the masses results, the Regge trajectories related to the pomeron and to the odderon in two different cases for the non-perturbative QCD beta functions known in the literature, namely, the beta function with a linear asymptotic behaviour and the beta function with an IR fixed point at finite coupling. At this point, we show that ASW model needs some corrections to provide compatible Regge trajectories related to the pomeron and to the odderon. In section \ref{DASW} we impose dynamical corrections coming from the dilaton field in the ASW model to achieve the Regge trajectories related to the pomeron and to the odderon. In particular, for the beta function with an IR fixed point at finite coupling, the Regge trajectories obtained for both even and odd spins glueball states, related to the pomeron and the odderon respectively,  are in agreement with those found in the literature. In  section \ref{Conc} we make some comments and summarise our results.


\section{The Anomalous Softwall Model}\label{ASW}

Let us start this section performing a quick review of the SW model. Then, we proceed to discuss the modifications due to the anomalous dimensions of the ASW model, introduce higher spin states and obtain glueball masses and Regge trajectories. 

The softwall model for scalar fields can be defined by the action \cite{Colangelo:2007pt,BoschiFilho:2012xr} 
\begin{equation}\label{acao_sw}
S = \frac{1}{\kappa} \int d^5 x \sqrt{-g} \; e^{-\Phi(z)} [ g^{MN} \partial_M X \partial_N X + M_5^2 X^2 ]
\end{equation}
\noindent where $\Phi(z)=kz^2$ is the dilatonic field,  $M_{5}$ is the 5-dimensional mass of the scalar field $X=X(z,x)$, and $g$ is the determinant of the metric tensor of the $5-$dimensional space AdS space described by:
\begin{equation}\label{ads}
ds^2= g_{MN} dx^M dx^N = \frac{R^2}{z^2} \, (dz^2 + \eta_{\mu \nu}dx^\mu dx^\nu)\,,
\end{equation}
\noindent with $M,N = 0,1,2,3,4\,; \; \mu, \nu = 0,1,2,3 \; {\rm and}\;  \eta_{\mu \nu} =$ diag $(-1, 1, 1, 1)$ is the metric of the four-dimensional Minkowski space.  The constant $R$ is called the radius of the $AdS_5$ space. 

In the SW model, to find the masses of glueball states one has to solve a ``Schr\"odinger-like'' equation:
\begin{equation}\label{sl}
- \psi''(z) + \left[ k^2 z^2 + \frac{15}{4z^2}  + 2k  +\left(\frac{R}{z} \right)^2 M^2_{5} \right]\psi(z) = (- q^2 )\psi(z)\;, 
\end{equation}
\noindent 
corresponding to the equation of motion for the field $X$, as shown in refs. \cite{Karch:2006pv,Colangelo:2007pt}. 
The glueball masses $m_n$, where $m_n^2 \equiv -q^2$,  
 can be obtained from normalisable solutions of Eq.(\ref{sl}) as 
 \begin{equation}
 m_n^2= \big[4n + 4 + 2 \sqrt{4+ R^2 M_5^2}\big]k \;,
 \end{equation}
 where $n=0,1,2,...$ denote the radial (spin 0) modes. 

From the AdS/CFT dictionary one can show how to relate masses of supergravity fields in the $AdS_{5} \times S^5$ space (here and in the following we are disregarding the $S^5$ space since the fields are assumed to be  independent of it) and scaling dimensions of the local gauge-invariant dual operator in the Super Yang-Mills theory (SYM). For instance, the classical (non-anomalous) conformal dimension $\Delta_{\rm{class.}}$ of a super Yang-Mills scalar operator is given by:
\begin{equation}\label{di}
\Delta_{\rm{class.}} = 2 + \sqrt{4 + R^2 M^2_{5}}\;. 
\end{equation}
Therefore, one can write:
\begin{equation}\label{dim}
R^2 M^2_{5} = \Delta_{\rm{class.}}( \Delta_{\rm{class.}} - 4) \;. 
\end{equation}

For a pure SYM theory defined on the boundary of the $AdS_5$ space, one has that the scalar glueball state $0^{++}$ is represented by the operator ${\cal O}_4$, given by:
\begin{equation}\label{fmn}
{\cal O}_4 = Tr(F^2) = Tr(F^{\mu\nu}F_{\mu \nu}).
\end{equation}
The lightest scalar glueball $0^{++}$ is dual to the fields with zero mass $(M^2_{5} = 0 )$ in the $AdS_5$ space, then one can see, using (\ref{dim}), that the operator ${\cal O}_4$ has conformal dimension $\Delta_{\rm{class.}} = 4$.

The SYM is a conformal theory, so the beta function vanishes and the conformal dimensions has no anomalous contributions, therefore they keep only their classical dimension. 

On the other side, for a QCD scalar glueball operator, its full dimension can be computed from the trace anomaly of the energy-momentum tensor \cite{Narison:1988ts,Gubser:2008yx}:
\begin{equation}\label{beta1}
T^{\mu}_{\mu} = \frac{\beta(\alpha)}{16 \pi \alpha^ 2} Tr F^2 + {\rm fermionic \;\;terms}
\end{equation}

\noindent where the beta function is defined as:
\begin{equation}\label{beta2}
\beta(\alpha(\mu) )\equiv \frac{d \alpha(\mu)}{d \ln(\mu)},
\end{equation}
and $\mu$ is a renormalisation scale, $\alpha \equiv g_{YM}^2 /4 \pi$ and $g_{YM}$ is the Yang-Mills coupling constant. 
The fermionic part in (\ref{beta1}) can be disregarded because only the operator $Tr F^2$ is relevant for our purposes. 
Moreover, the scaling behaviour for a generic operator is given by:
\begin{equation}\label{beta3}
\Delta_{\cal O} = - \frac{d {\cal O}}{d \ln \mu}.
\end{equation}
The full dimension $\Delta_{\cal O}$ also can be seen as a sum of the classical dimension $\Delta_{\rm{class.}}$ and the anomalous dimension $\gamma(\mu)$, so that:
\begin{equation}\label{beta4}
\Delta_{\cal O} = \Delta_{\rm{class.}} + \gamma(\mu).
\end{equation}

For the scalar glueball operator, one can insert eq.(\ref{beta1}), without fermionic part,  in (\ref{beta3}), obtaining:
\begin{equation}\label{beta5}
\Delta_{T^{\mu}_{\mu}} \left( \frac{\beta(\alpha)}{8 \pi \alpha^ 2} Tr F^2 \right)  = - \frac{d}{d \ln \mu} \left( \frac{\beta(\alpha)}{8 \pi \alpha^ 2} Tr F^2 \right)\;,
\end{equation}

\noindent or
\begin{equation}
\Delta_{T^{\mu}_{\mu}} \left( \frac{\beta(\alpha)}{8 \pi \alpha^ 2} Tr F^2 \right)  = - (\beta'(\alpha) - \frac{2}{\alpha} \beta(\alpha) - \Delta_{F^2}) \frac{\beta(\alpha)}{8 \pi \alpha^ 2} Tr F^2\;,
\end{equation}

\noindent where the prime represents the derivative with respect to $\alpha$.

The trace $T^\mu_\mu$ scales classically, that means $\Delta_{T^{\mu}_{\mu}\,(Class.)} = 4$ \cite{Gubser:2008yx}. So, the scalar glueball operator $Tr F^2$ has the full dimension:
\begin{equation}\label{beta6}
\Delta_{F^2} = 4 + \beta'(\alpha) - \frac{2}{\alpha} \beta(\alpha)
\end{equation}

Using the 't Hooft coupling $\lambda \equiv N_C g_{YM}^2 = 4 \pi N_C \alpha$, one gets
\begin{equation}\label{beta7}
\Delta_{F^2} = 4 + \beta'(\lambda) - \frac{2}{\lambda} \beta(\lambda)
\end{equation}

\noindent and now the prime represents the derivative with respect to $\lambda$ and the beta function is given by:
\begin{equation}
\beta(\lambda(\mu)) \equiv \frac{d \lambda(\mu)}{d \ln(\mu)}.
\end{equation}

In the reference \cite{BoschiFilho:2012xr} this approach was used to study the scalar glueball and its radial (spin 0) excitations spectroscopy due to contributions of anomalous dimensions.

In this present work our concern is about higher spin glueballs, for even and odd spins, and to get their corresponding Regge trajectories related to the pomeron and to the odderon.

To get the Regge trajectory for even glueball states we will use the method described in \cite{deTeramond:2005su} to raise the spin of the glueball, that is, we will insert  symmetrised covariant derivatives in a given operator with spin $S$ in order to raise the total angular momentum, such that, the total angular momentum after the insertion is now $S+J$. 
In the particular case of the operator ${\cal O}_4 = F^2$, one gets:
\begin{equation}\label{4+J}
{\cal O}_{4 + J} = FD_{\lbrace\mu1 \cdots} D_{\mu J \rbrace}F,
\end{equation}
\noindent with conformal dimension $\Delta_{\rm{class.}} = 4 + J$ and spin $J$. The reference \cite{BoschiFilho:2005yh} used this approach within the hardwall model to calculate the masses of glueball states $0^{++}$, $2^{++}$, $4^{++}$, etc and to obtain the Regge trajectory for the pomeron in agreement with those found in the literature. 

So, for even spin glueball states using the softwall model after the insertion of symmetrised covariant derivatives, and using that $\Delta_{\rm{class.}} = 4 + J $ and $\Delta_{\rm{class.}} = 2 + \sqrt{4 + R^2 M^2_{5}}$, one has:
\begin{equation}\label{hsp}
R^2M^2_{5} = J(J+4)\,; \qquad ({\rm even}\, J)\;. 
\end{equation}
In a similar way,  one can write the full dimension $\Delta^{even\,J}_{T^{\mu}_{\mu}} = 4 + J$, 
and now the eq. (\ref{beta7}) can be written as:
\begin{equation}\label{beta8}
\Delta^{even\, J}_{F^2} = 4 + J + \beta'(\lambda) - \frac{2}{\lambda} \beta(\lambda).
\end{equation}
Using (\ref{dim}), the full dimension for a glueball state with higher even spin $J$, taking into account the beta function is:
\begin{equation}\label{dfullp}
R^2M^2_{5} = \Delta^{even\, J}_{F^2}  (\Delta^{even\, J}_{F^2}  -4)
\end{equation}
or explicitly:
\begin{equation}\label{r2par}
R^2M^2_{5} = \left[  4 + J + \beta'(\lambda) - \frac{2}{\lambda} \beta(\lambda)\right] \left[ J + \beta'(\lambda) - \frac{2}{\lambda} \beta(\lambda)\right]\,; \qquad ({\rm even}\, J)\,.
\end{equation}

On the other hand, for odd spin glueballs, the operator ${\cal O}_6$ that describes the glueball state $1^{--}$ is given by:
\begin{equation}
 {\cal O}_{6} =SymTr\left( {\tilde{F}_{\mu \nu}}F^2\right),
 \end{equation} 

\noindent and after the insertion of symmetrised covariant derivatives one gets:
\begin{equation}\label{6+J}
{\cal O}_{6 + J} = SymTr\left( {\tilde{F}_{\mu \nu}}F D_{\lbrace\mu1 \cdots} D_{\mu J \rbrace}F\right),
\end{equation}

\noindent with conformal dimension $\Delta_{\rm{class.}} = 6 + J$ and spin $1+J$. The reference \cite{Capossoli:2013kb} used this approach within the hardwall model to calculate the masses of odd spin glueball states $1^{--}$, $3^{--}$, $5^{--}$, etc and the Regge trajectory for the odderon were obtained in agreement with those found in the literature.

For the case of the odd spin glueball states, using $\Delta = J + 6$ and $\Delta = 2 + \sqrt{4 + R^2 M^2_{5}}$\,, one has:
\begin{equation}\label{hsi}
R^2 M^2_{5} = (J+6)(J+2)\,; \qquad ({\rm odd}\, J).
\end{equation}

In a similar way, one can write the full dimension $\Delta^{odd\, J}_{T^{\mu}_{\mu}} = 6 + J$\,, 
and now the eq. (\ref{beta7}) can be written as:
\begin{equation}\label{beta8_1}
\Delta^{odd\, J}_{F^2} = 6 + J + \beta'(\lambda) - \frac{2}{\lambda} \beta(\lambda).
\end{equation}
Using (\ref{dim}), one can write the full dimension for a glueball state with higher odd spin $J$, taking into account the beta function:
\begin{equation}\label{dfulli}
R^2M^2_{5} = \Delta^{odd\, J}_{F^2}  (\Delta^{odd\, J}_{F^2}  -4)
\end{equation}
and explicitly:
\begin{equation}\label{r2impar}
R^2M^2_{5} = \left[  6 + J + \beta'(\lambda) - \frac{2}{\lambda} \beta(\lambda)\right] \left[ 2 + J + \beta'(\lambda) - \frac{2}{\lambda} \beta(\lambda)\right]\,; \qquad ({\rm odd}\, J).
\end{equation}

Then, replacing eqs. (\ref{r2par}) and (\ref{r2impar}) in the Schr\"odinger-like equation (\ref{sl}) one can solve it numerically and get the masses of higher spin glueballs (even and odd) and consequently obtain the Regge trajectories for the pomeron and the odderon, as we will show below. 

At this point, we make a brief comment about QCD beta functions. From perturbative QCD it is well known that one can express the beta function through a power series of the coupling where each term comes from a certain loop order. Exceptionally the two first terms do not depend on the renomalisation set up, but other ones, i.e, higher order terms do.

For our purposes, we will consider some effective non-perturbative beta functions that could reproduce the IR behaviour of QCD as one can see in the works \cite{Zeng:2008sx, Alanen:2009na, Ryttov:2007cx}. 
Another requirement is that the beta functions reproduce the ultraviolet perturbative behaviour analogous to the QCD for small $\lambda$ in $1-$ loop approximation. That is:
$ \beta(\lambda) \sim - b_0 \lambda^2$\;,
 where $b_0$ is a universal coefficient of the perturbative QCD beta function at leading order, given by 
$ b_0 = \frac{1}{8 \pi^2}\left( \frac{11}{3} - \frac{2}{9} N_f\right) $\;. 
For a pure $SU(3)_c$ one has $N_f = 0$, then $b_0 = 11/24 \pi^2$.

From the AdS/QCD softwall model one can relate the holographic or radial coordinate $z$ of the $AdS_5$ space with $\mu^{-1}$ where $\mu$ was defined as the renormalisation group scale. So, the relation of the beta function and $z$ is then:
\begin{equation}\label{beta10}
\beta(\lambda(\mu)) = \mu \frac{d \lambda(\mu)}{d \mu} \Rightarrow  \beta(\lambda(z)) = - z \frac{d \lambda(z)}{dz},
\end{equation}

\noindent where the integration constant will be fixed by $\lambda(z) \equiv \lambda_0$ at a particular energy scale $z_0$.

In the following subsections the ``Schr\"odinger-like'' equation (\ref{sl}) will be solved numerically for two regimes, namely, the beta function with a linear asymptotic behaviour and the beta function with an IR fixed point at finite coupling. Then, we obtain the corresponding Regge trajectories trying to fit the pomeron and the odderon for each beta function. 


\subsection{Beta function with a linear IR asymptotic behaviour}

We begin considering the beta function given in \cite{Zeng:2008sx, Ryttov:2007cx}, such that:
\begin{equation}\label{beta13}
\beta(\lambda) = - \frac{ b_0 \lambda^2}{1 + b_1 \lambda}  \;;\;\;\; {\rm for}\;\;\; b_0, b_1 > 0\;.
\end{equation}
This beta function behaves as  $ - b_0 \lambda^2$ in the UV and as    $ - (b_0/b_1) \lambda$  in the IR.  
Solving eq.~(\ref{beta10}) for this beta function, one gets exactly:
\begin{equation}\label{beta14}
\lambda(z) = \frac{1}{b_1 W\left( \frac{\exp^{\frac{1}{b_1 \lambda_0}}}{b_1 \lambda_0}(\frac{z_0}{z})^{b_0/b_1}\right) }\;,
\end{equation}
where $W(z)$ is the Lambert function. 

Substituting eqs. (\ref{beta13}) and (\ref{beta14}) in eqs. (\ref{r2par}) and (\ref{r2impar}), solving numerically the Schr\"odinger-like equation (\ref{sl}) and using some sets of values for $k$, $b_1$ and $\lambda_0$, one can get the Regge trajectories for even and odd glueball states, that could be related to the pomeron and to the odderon, respectively.

\begin{table}[h]

\vspace{0.5 cm}
\centering
\begin{tabular}{|c|c|c|c|c|c|}
\hline\hline
Set & $k$ & $b_1\times 10^{3}$ &\; $\lambda_0$ \;& pomeron & odderon \\ 
\hline \hline
1& $- 0.25$ & $1.2 $ & $19$ & $J \approx (- 0.3 \pm 0.3) + (0.42 \pm 0.03) m^2$ & $J \approx (- 1.8 \pm 0.5) + (0.40 \pm 0.03) m^2$  \\ 
\hline 
2 & $-0.49$ & $1.2 $ & $19$ & $J \approx (- 0.3 \pm 0.4) + (0.39 \pm 0.03) m^2$ & $J \approx (- 1.3 \pm 0.4) + (0.34 \pm 0.02) m^2$ \\ 
\hline 
3 & $-0.72$ & $1.2 $ & $19$ & $J \approx (- 0.6 \pm 0.4) + (0.38 \pm 0.02) m^2$ & $J \approx (-1.6 \pm 0.3) + (0.33 \pm 0.01) m^2$ \\ 
\hline 
4 & $- 1.00$ & $1.2 $ & $19$& $J \approx (- 1.0  \pm 0.3) + (0.35 \pm 0.01) m^2$ & $J\approx (- 2.2 \pm 0.3) + (0.32 \pm 0.01) m^2$ \\ 
\hline 
5 & $- 1.00$ & $1.2 $ & $16$ & $J \approx (- 1.7  \pm 0.2) + (0.45 \pm 0.01) m^2$ & $J \approx (- 3.2 \pm 0.2) + (0.43 \pm 0.01) m^2$ \\ 
\hline 
6 & $- 1.00$ & $1.0 $ & $16$& $J \approx (- 1.7 \pm 0.2) + (0.45 \pm 0.01) m^2$ & $J\approx (- 3.2 \pm 0.2) + (0.42 \pm 0.01) m^2$ \\ 
\hline 
\end{tabular}
\caption{\em Different values of  $k$, $b_1$ and $\lambda_0$ used in the ASW model for the beta function with a linear IR asymptotic behaviour, eq. \eqref{beta13}, and the results for the Regge trajectories obtained for the pomeron and the odderon. The errors come from linear fit.}
\label{t4}
\end{table}

The values for $k$, $b_1$ and $\lambda_0$, and the results for the Regge trajectories are presented in table \ref{t4}, where one can see that the Regge trajectories found for both the pomeron and the odderon for the beta function with a linear IR asymptotic behaviour are in disagreement with those found in  \cite{Landshoff:2001pp}, for the pomeron, and in \cite{LlanesEstrada:2005jf} for the odderon. For instance, in the 4th set of table \ref{t4}, one finds that the angular coefficients found for the pomeron and for the odderon are respectively  $0.35\pm 0.01$  and $0.32 \pm 0.01$ GeV$^{-2}$, which are much higher than the expect values of $0.25$ GeV$^{-2}$ for the pomeron, eq. \eqref{pomeron}, and $0.23$ or $0.18$ GeV$^{-2}$ for the odderon, eqs. (\ref{r_odderon}) and  (\ref{nr_odderon}). Since for the other sets the values for the angular coefficients are even higher, we conclude that the ASW model with the beta function with linear asymptotic behaviour does not give good results for the Regge trajectories for the pomeron or for the odderon.


\subsection{Beta function with an IR fixed point at finite coupling}

Here, we consider the beta function given in \cite{Alanen:2009na}:
\begin{equation}\label{beta11}
\beta(\lambda) = - b_0 \lambda^2 \left[ 1 - \frac{\lambda}{\lambda_{\ast}}\right] \;;\;\;\; {\rm for}\;\;\; \lambda_{\ast} > 0\;.
\end{equation}
This the beta function has the necessary IR and UV requirements. This means that for the IR fixed point $\lambda = \lambda_{\ast}$ this beta function vanishes. Moreover, it reproduces the perturbative $\beta(\lambda) \sim - b_0 \lambda^2$ at $1-$ loop order in the ultraviolet and behaves as $\beta (\lambda) \sim + \lambda^3$ at large coupling. 
The equation (\ref{beta10})  can also be solved exactly for this beta function, so that:
\begin{equation}\label{beta12}
\lambda(z) = \frac{\lambda_{\ast}}{1 + W\left(\left( \frac{z_0}{z}\right)^{b_0 \lambda_{\ast}} \left( \frac{\lambda_{\ast} - \lambda_0}{\lambda_0}\right)  \exp^{\frac{\lambda_{\ast} - \lambda_0}{\lambda_0}}\right) }
\end{equation}

\noindent where $W(z)$ is again the Lambert function and $\lambda(z_0) = \lambda_0$ fixes the integration constant. 
This equation leads to the expected QCD asymptotic behaviour at short distances when $z$ is close to the boundary $(z\to 0)$:
\begin{equation}
\lambda(z) \sim - 1/(b_0 \ln z).
\end{equation}

Replacing (\ref{beta11}) and (\ref{beta12}) in eqs. (\ref{r2par}) and (\ref{r2impar}), solving numerically the Schr\"odinger-like equation (\ref{sl})  and using some sets of values for $k$, $\lambda_0$ and $\lambda_{\ast}$, one can get Regge trajectories for even and odd glueball states, that could be related to the pomeron and odderon, respectively. 
The results obtained are presented in table \ref{t2}. 

\begin{table}[h]

\vspace{0.5 cm}
\centering
\begin{tabular}{|c|c|c|c|c|}
\hline
 $k$ & $\lambda_0$ & pomeron & odderon \\ 
\hline 
$\; - 0.36\; $ & \; $18.5$ \; & \;$J \approx (- 0.9 \pm 0.4) + (0.51 \pm 0.03) m^2$ \;&\; $J \approx (- 3.0 \pm 0.9) + (0.53 \pm 0.06) m^2$ \; \\ 
\hline 
$-0.16$ & $18.5$ & $J \approx (- 4 \pm 2) + (1.4 \pm 0.3) m^2$ & $J \approx (- 19.7 \pm 0.3) + (3.06 \pm 0.04) m^2$ \\ 
\hline 
$-0.36$ & $10.5$ & $J \approx (- 1.93 \pm 0.03) + (1.34 \pm 0.01) m^2$ & $J \approx (- 3.82 \pm 0.05) + (1.32 \pm 0.01) m^2$ \\ 
\hline 
$- 0.36$ & $25.5$  & $J \approx (- 3  \pm 2) + (0.57 \pm 0.10) m^2$ & $J \approx (- 14 \pm 2) + (1.0 \pm 0.1) m^2$ \\ 
\hline 
\end{tabular}
\caption{\em Regge trajectories obtained for both pomeron and odderon from ASW model using the beta function with an IR fixed point at finite coupling, eq. \eqref{beta11}, and $\lambda_\ast=350$. The errors come from linear fit.}
\label{t2}
\end{table}

From the table \ref{t2} one can see that the Regge trajectories found for both the pomeron and the odderon for the beta function with an IR fixed point at finite coupling regime are in disagreement with those found in  \cite{Landshoff:2001pp}, for the pomeron, and in \cite{LlanesEstrada:2005jf} for the odderon. As in the case of the previous beta function, for instance, the angular coefficients found here are too high when compared with the ones from the pomeron $0.25$, eq. \eqref{pomeron}, and the odderon  $0.23$ or $0.18$ GeV$^{-2}$, eqs. (\ref{r_odderon}) and  (\ref{nr_odderon}). So, we conclude that the ASW model with the beta function with an IR fixed point at finite coupling does not give reasonable results for the Regge trajectories for the pomeron or the odderon. 



\section{The Dynamical Corrections to the Anomalous Softwall Model}\label{DASW}

In this section we will apply dynamical corrections to modify the ASW model to investigate if these corrections can provide Regge trajectories for the pomeron and the odderon compatible with those found in the literature. To do this, let us  perform a quick review of the dynamical softwall  (DSW) model, discussed in \cite{Shock:2006gt, White:2007tu, Li:2013oda}. 

The $5D$ action for the graviton-dilaton coupling in the string frame is given by:
\begin{equation}\label{acao_corda}
S = \frac{1}{16 \pi G_5} \int d^5 x \sqrt{-g_s} \; e^{-2\Phi(z)} (R_s + 4 \partial_M \Phi \partial^M \Phi - V^s_G(\Phi))
\end{equation}
\noindent where $G_5$ is the Newton's constant in five dimensions, $g_s$ is the metric tensor in the $5-$dimensional space, $\Phi=\Phi(z)$ is the dilaton field and $V_G$ is the dilatonic potential. All of these parameters are defined in the string frame. 
The metric $g_s$ has the following form:
\begin{equation}\label{g_s}
ds^2 = g^s_{MN} dx^M dx^N = b^2_s(z)(dz^2 + \eta_{\mu \nu}dx^\mu dx^\nu); \; \; \;b_s(z) \equiv e^{A_s(z)}
\end{equation}
\noindent with $M,N = 0,1,2,3,4; \; \mu, \nu = 0,1,2,3,$ and  $\eta_{\mu \nu} =$ diag $(-1, 1, 1, 1)$ is the metric of the four-dimensional Minkowski space. 
Performing a Weyl rescaling, from the string frame to the Einstein frame, one can write eq. (\ref{acao_corda}) as:
\begin{equation}\label{acao_einstein}
S = \frac{1}{16 \pi G_5} \int d^5 x \sqrt{-g_E} \; (R_E -\frac{4}{3} \partial_M \Phi \partial^M \Phi - V^E_G(\Phi))\;, 
\end{equation}
\noindent where
\begin{equation}\label{weyl}
 g^E_{MN} = g^s_{MN}e^{-\frac{2}{3}\Phi}\;; \qquad V^E_G = e^{\frac{4}{3}\Phi}V^s_G\;.
\end{equation}
Varying the action (\ref{acao_einstein}), one can obtain the equations of motion, which are given by:
\begin{equation}\label{eq_mov_e_1}
 E_{MN}  + \frac{1}{2}g^E_{MN}\left(\frac{4}{3} \partial_L\Phi \partial^L\Phi + V^E_G \right) - \frac{4}{3} \partial_M \Phi \partial^M \Phi = 0\;;
\end{equation}
\begin{equation}\label{eq_mov_e_2}
 \frac{8}{3 \sqrt{g_E}} \partial_M (\sqrt{g_E} \partial^M \Phi) - \partial_\Phi V^E_G(\Phi)  = 0\;,
\end{equation}
\noindent where $E_{MN}$ is the Einstein tensor.

Using the metric parametrisation given by (\ref{g_s}), the equations of the motion (\ref{eq_mov_e_1}) and (\ref{eq_mov_e_2}) can be written as:
\begin{equation}\label{eq_mov_e_2_1}
 -A''_E + A'^2_E - \frac{4}{9}\Phi'^2  = 0\;;
\end{equation}
\begin{equation}\label{eq_mov_e_2_2}
 \Phi'' + 3A'_E \Phi' - \frac{3}{8}e^{2A_E}\partial_\Phi V^E_G(\Phi) = 0\;,
\end{equation}
\noindent where we defined:
\begin{equation}\label{redef}
b_E (z) = b_s(z)e^{-\frac{2}{3}\Phi(z)} = e^{A_E(z)}\;; \qquad A_E(z) = A_s(z) - \frac{2}{3}\Phi(z)\;.
\end{equation}
\noindent Solving the equations (\ref{eq_mov_e_2_1}) and (\ref{eq_mov_e_2_2}) for the quadratic dilaton background, $\Phi(z)=kz^2$, one finds:
\begin{equation}\label{sol_eq_mov_e_2_1}
 A_E(z) = \log{\left( \frac{R}{z} \right)} - \log{\left(_0F_1\left(\frac 54, \frac{\Phi^2}{9}\right)\right)}\;, 
\end{equation}
\noindent and
\begin{equation}\label{sol_eq_mov_e_2_2}
 V^E_G(\Phi) = -\frac{12 ~ _0F_1(\frac14, \frac{\Phi^2}{9})^2}{R^2} + \frac{16 ~ _0F_1(\frac 54, \frac{\Phi^2}{9})^2\, \Phi^2}{3 R^2}\;,
\end{equation}
where $_0F_1(a,z)$ is a confluent hypergeometric function. 
Using (\ref{redef}) and (\ref{sol_eq_mov_e_2_1}), one can easily see that:
\begin{equation}\label{redef_2}
 A_s(z) = \log{\left( \frac{R}{z} \right)}  + \frac{2}{3}\Phi(z) - \log{\left(_0F_1\left(\frac 54, \frac{\Phi^2}{9}\right)\right)}\;, 
\end{equation}
which implies that the metric \eqref{g_s} in this dynamical model is no longer AdS, but it is asymptotically AdS in the limit $z \to 0$. 

The $5D$ action for the scalar glueball in the string frame is given by  \cite{Colangelo:2007pt, Forkel:2007tz}:
\begin{equation}\label{acao_ori_soft}
S = \int d^5 x \sqrt{-g_s} \; \frac{1}{2} e^{-\Phi(z)} [\partial_M {\cal G}\partial^M {\cal G} + M^2_{5} {\cal G}^2]\;,
\end{equation}
\noindent and its equations of motion are:
\begin{equation}\label{eom_1}
\partial_M[\sqrt{-g_s} \;  e^{-\Phi(z)} g^{MN} \partial_N {\cal G}] - \sqrt{-g_s} e^{-\Phi(z)} M^2_{5} {\cal G} = 0\;. 
\end{equation}
\noindent Using the metric (\ref{g_s}), one gets: 
\begin{equation}\label{eom_3}
\partial_M[e^{3A_s(z) - \Phi(z)} ~ \eta^{MN}  \partial_N {\cal G}] - e^{5 A_s(z)-\Phi(z)} M^2_{5} {\cal G} = 0\;.
\end{equation}
\noindent Doing the substitution 
$B(z) = \Phi(z) - 3A_s(z)$, 
using the ansatz 
${\cal G}(z, x^{\mu}) = v(z) e^{i q_{\mu} x^{\mu}}$, 
and defining $v(z) = \psi (z) e^{\frac{B(z)}{2}}$, one gets: 
\begin{equation}\label{equ_5}
- \psi''(z) + \left[ \frac{B'^2(z)}{4}  - \frac{B''(z)}{2} +  \left( \frac{R M_{5}}{z}\right)^2  e^{4kz^2/3} {\cal A}^{-2} \right] \psi(z) = - q^2 \psi(z) \;, 
\end{equation}
where ${\cal A}$ is given by $_0F_1(\frac 54, \frac{\Phi^2}{9})$, or explicitly, for the quadratic dilaton $\Phi(z)= k z^2$: 
\begin{equation}\label{equ_7_new1}
- \psi''(z) + \left[ k^2 z^2 + \frac{15}{4z^2}  - 2k +  \left( \frac{R M_{5}}{z}\right)^2  e^{4kz^2/3} {\cal A}^{-2} \right] \psi(z) = - q^2 \psi(z) \;. 
\end{equation}
This equation was solved numerically in \cite{Li:2013oda} and the masses found for the scalar glueball and its radial (spin 0) excitations are  compatible with those obtained by lattice QCD.

To take into account dynamical corrections plus the anomalous dimension effects, one must recall, from section $2$, that for even spin glueball states the full dimension is given by eq.(\ref{beta8}), so that:
\begin{equation}
\Delta^{even\, J}_{F^2} = 4 + J + \beta'(\lambda) - \frac{2}{\lambda} \beta(\lambda).
\end{equation}

For the odd spin glueball state the full dimension is given by eq.(\ref{beta8_1}), then
\begin{equation}
\Delta^{odd\, J}_{F^2} = 6 + J + \beta'(\lambda) - \frac{2}{\lambda} \beta(\lambda).
\end{equation}

Using once again the relation $R^2 M^2_5 = \Delta (\Delta - 4)$, one gets:
\begin{equation}\label{mass}
R^2 M^2_5 = \left\{
\begin{array}{rcl}
\left[  4 + J + \beta'(\lambda) - \frac{2}{\lambda} \beta(\lambda)\right] \left[ J + \beta'(\lambda) - \frac{2}{\lambda} \beta(\lambda)\right], & {\rm if} & \Delta = \Delta^{even\, J}_{F^2}  \\
\left[  6 + J + \beta'(\lambda) - \frac{2}{\lambda} \beta(\lambda)\right] \left[ 2 + J + \beta'(\lambda) - \frac{2}{\lambda} \beta(\lambda)\right], & {\rm if} & \Delta = \Delta^{odd\, J}_{F^2}
\end{array} \right.
\end{equation}

To consider even or odd spin glueball states one can replace the first or the second line of eq. (\ref{mass}), respectively,  into the Schr\"odinger-like equation obtained from the DSW softwall model, eq.(\ref{equ_7_new1}),  and solve it numerically.  For our purposes, the same two regimes of the previous section will be studied, namely, the beta function with a linear asymptotic behaviour and the beta function with an IR fixed point at finite coupling. The results will be discussed in the following.

\subsection{Beta function with a linear IR asymptotic behaviour}

Using the beta function given by eq. (\ref{beta13})  and replacing it into eq. \eqref{mass}
 one can solve numerically the Schr\"odinger-like equation for the DSW model, (\ref{equ_7_new1}), for both even and odd glueball states.

For a set of values for $k$, $b_1$ and $\lambda_0$, one can get Regge trajectories for even and odd glueball states. The set used together with the results obtained for Regge trajectories for both the pomeron and the odderon  are shown in table \ref{t7}. 

\begin{table}[h]

\vspace{0.5 cm}
\centering
\begin{tabular}{|c|c|c|c|c|}
\hline
 $k$ & $b_1 \times 10^{3}$ &\; $\lambda_0 \;$  & pomeron & odderon \\ 
\hline 
$- 0.25$ & $1.2$ & $19$& $J \approx (- 14 \pm 1) + (4.1 \pm 0.3) m^2$ & $J \approx (- 21 \pm 1) + (5.0 \pm 0.3) m^2$  \\ 
\hline 
\end{tabular}
\caption{\em Parameters used in the beta function with a linear IR asymptotic behaviour, eq. (\ref{beta13}),  and the corresponding Regge trajectories obtained for both the pomeron and the odderon from the anomalous and dynamical softwall model, eqs. (\ref{equ_7_new1}) and \eqref{mass}. The errors come from linear fit.}
\label{t7}
\end{table}

From the table \ref{t7},  one can see that the Regge trajectories found for both the pomeron and the odderon with the beta function with a linear IR asymptotic behaviour, eq. (\ref{beta13}), are in disagreement with those found in  \eqref{pomeron} for the pomeron  \cite{Landshoff:2001pp}, and in eqs. (\ref{r_odderon}) and  (\ref{nr_odderon})  for the odderon \cite{LlanesEstrada:2005jf}. The angular coefficients are too high and the intercepts are too low, both for the pomeron and for the odderon. Other sets of parameters give even poorer results.


\subsection{Beta Function with an IR fixed point at finite coupling}

Using the beta function given by eq. (\ref{beta11}) and substituting it into 
 eq. \eqref{mass}, one can solve numerically the Schr\"odinger-like equation (\ref{equ_7_new1}) for both even and odd glueball states.

We consider different sets of values for $k$, $\lambda_0$ and $\lambda_{\ast}$, and get the masses of the glueball states with even and odd spins and the Regge trajectories related to the pomeron and odderon, respectively. The sets of parameters used and the results obtained for the glueball state masses with even and odd spins are presented in the table  \ref{t19}. 
The results obtained for Regge trajectories are shown in table~ \ref{t9}.

\begin{table}[!h]
\centering
\begin{tabular}{|c||c|c|c||c|c|c|c|c|c||c|c|c|c|c|c|}
\hline  & \multicolumn{3}{|c||}{Parameters}
 &  \multicolumn{12}{c|}{Glueball States $J^{PC}$}  \\  
\cline{2-16}
Set& $k$ & $\lambda_0$ & $\lambda_{\ast}$  & $0^{++}$ & $2^{++} $ & $4^{++}$ & $6^{++}$ & $8^{++}$ & $10^{++}$  &  $1^{--}$ & $3^{--} $ & $5^{--}$ & $7^{--}$ & $9^{--}$ & $11^{--}$  \\
\hline \hline                        
 1& $ 0.16$ & $18.5$ & $350$ & 1.69 & 3.28 & 4.76 & 6.23  & 7.67  & 9.12    & 4.02 & 5.50 & 6.95 & 8.40  & 9.84   & 10.00  \\ \hline                              
 2& $ 0.09$ & $18.5$ & $350$ & 1.62 & 2.84  & 4.00 &  5.14  & 6.26  &  7.37   & 3.42 & 4.57 & 5.70 &  6.82  & 7.93 & 9.04   \\ \hline                                 
 3& $ 0.04$ & $18.5$ & $350$ & 1.56 & 2.52 & 3.43 &  4.32  & 5.19  & 6.05  & 2.98 & 3.88 & 4.76  &  5.62  & 6.48  & 7.32  \\ \hline                             
 4& $ 0.09$ & $10.5$ & $350$  & 0.79 & 2.13 & 3.28 &  4.39  & 5.48  & 6.57  & 2.72 & 3.84 & 4.94 & 6.03 & 7.11 & 8.19  \\ \hline                             
 5& $0.09$ & $18.5$ & $250$ & 1.64 & 2.86 & 4.02 & 5.16 & 6.28 & 7.39  & 3.44  & 4.59 & 5.72 &  6.84  & 7.95  & 9.05  \\ \hline 
\hline
 \end{tabular} 
\caption{\em  Masses {\rm (GeV)}  for the glueball states $J^{PC}$ with even and odd $J$ with $P=C=\pm 1$ calculated from the anomalous dynamical softwall model, eqs. \eqref{equ_7_new1} and \eqref{mass}, and the beta function with an IR fixed point at finite coupling, \eqref{beta11}, using five sets of parameters  $k$ {\rm (GeV$^{2}$)},  $\lambda_0$ and $\lambda_{\ast}$ (dimensionless).}
\label{t19}
\end{table}

\begin{table}[h]

\vspace{0.5 cm}
\centering
\begin{tabular}{|c|c|c|}
\hline
 Set & pomeron & odderon \\ 
\hline 
 $1$ & $\; J\approx (0.6 \pm 0.5) + (0.12 \pm 0.01) m^2$ & $\; J \approx (- 0.1 \pm 0.4) + (0.10 \pm 0.01) m^2$  \\ 
\hline 
 $2$ & $J \approx (0.4 \pm 0.5) + (0.19 \pm 0.02) m^2$ & $J \approx (- 0.4 \pm 0.4) + (0.15 \pm 0.01) m^2$ \\ 
\hline 
 $3$ & $J\approx (0.1 \pm 0.5) + (0.28 \pm 0.02) m^2$ & $J \approx (- 0.8 \pm 0.4) + (0.24 \pm 0.02) m^2$ \\ 
\hline 
 $4$ & $J \approx (0.9  \pm 0.5) + (0.23 \pm 0.02) m^2$ & $ J \approx (0.1 \pm 0.4) + (0.18 \pm 0.01) m^2$ \\ 
\hline 
 $5$ & $J \approx (0.4  \pm 0.5) + (0.19 \pm 0.02) m^2$ & $J  \approx (- 0.4 \pm 0.4) + (0.15 \pm 0.01) m^2$ \\ 
\hline 
\end{tabular}
\caption{\em Regge trajectories obtained for both pomeron and odderon from the anomalous softwall model with dynamical corrections, eqs. (\ref{equ_7_new1}) and \eqref{mass}, using the beta function with an IR fixed point at finite coupling, \eqref{beta11},  for the sets of parameters presented in table \ref{t19}. The errors come from linear fit.}
\label{t9}
\end{table}
\begin{figure}[h] 
  \centering
  \includegraphics[scale = 0.5]{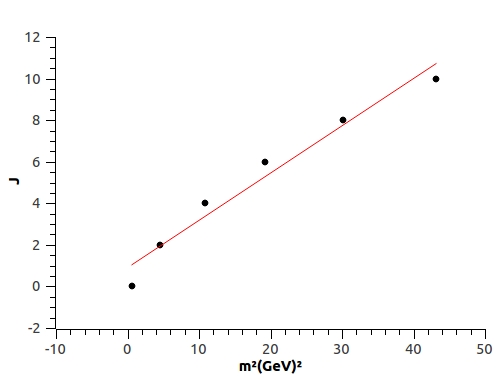}\qquad \includegraphics[scale = 0.5]{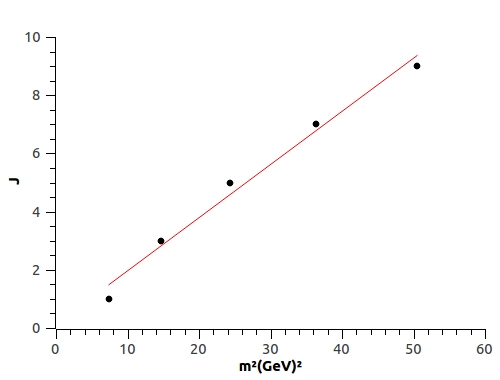} 
\caption{\em Approximate Regge trajectories for the pomeron (left) and odderon (right) using data from table \ref{t19} with set 4. The pomeron Regge trajectory corresponds to the states $0^{++}$, $\dots$, $10^{++}$, and the odderon Regge trajectory to the states $1^{--}$, $\dots$, $9^{--}$.}
\label{plot5}
\end{figure}

From the table \ref{t9}, one can see that the Regge trajectories found, using set 4 presented in table \ref{t19}, for both pomeron and odderon for the beta function with an IR fixed point at finite coupling regime are in agreement with those found in \eqref{pomeron}  for the pomeron  \cite{Landshoff:2001pp}, and with the non-relativistic constituent model, eq. \eqref{nr_odderon}, for the odderon \cite{LlanesEstrada:2005jf}.
In particular, the Regge trajectory related to the pomeron from table \ref{t9} with set 4, is given by:
\begin{equation}\label{pad}
J(m^2) \approx  (0.9  \pm 0.5) + (0.23 \pm 0.02) \, m^2 \,,
\end{equation}
 \noindent and is shown in the left panel of figure  \ref{plot5}.  
The Regge trajectory related to the odderon from table \ref{t9} with set 4, excluding the glueball state $11^{--}$, is given by:
\begin{equation}\label{oad}
J(m^2) \approx   (0.1 \pm 0.4) + (0.18 \pm 0.01) m^2\,,
\end{equation}
 \noindent  which is shown in the right panel of figure  \ref{plot5}.  Here, we excluded the state $11^{--}$ because the corresponding Regge trajectory with it was not good when compared with the odderon results, eqs. \eqref{r_odderon} and \eqref{nr_odderon}. Note that this happens due to the non-linear behaviour of the quadratic masses of the odd spin glueballs which can be computed by using the values for the masses found in table \ref{t19}. 
 Note also that in the original softwall model the Regge trajectories are linear, but here in the dynamical version of this model the trajectories are no longer linear. As we are going to show on the next section, this happens because the dynamical corrections produce effective potentials which turns this model similar to the hardwall model which provides non-linear trajectories. These non-linear behaviour also lead to large uncertainties for the intercepts. As mentioned before, those uncertainties come from the usual linear (least squares) regression fit.


\section{Discussion and conclusions}\label{Conc}

\begin{figure}[!ht] 
  \centering
\quad  \includegraphics[scale = 0.60]{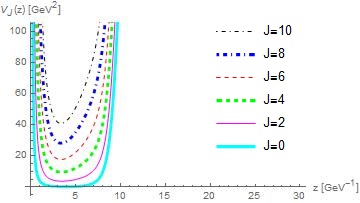} \quad \includegraphics[scale = 0.60]{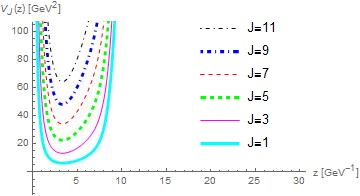} 
\caption{\em Plots for the effective potential  $V_J(z)$ of the dynamical anomalous softwall model against the holographic coordinate $z$ for even spins  $J=0, \dots, 10$ $(left)$, and for odd spins $J=1,  \dots, 11$ $(right)$.}
\label{plot9}
\end{figure}

In this work we used the anomalous dimension softwall model, related to QCD beta functions to obtain high spins glueball masses and the corresponding Regge trajectories. Then, we take into account the dynamical corrections caused by the dilaton field, such that the model become a solution of Einstein's equations in five dimensions. We take this anomalous and dynamical model to calculate the masses of glueball states with even and odd spins. 

Our motivation to consider these dynamical corrections, as was shown in the section II, is due to the fact that although the ASW model worked well for scalar glueball states and its radial (spin 0) excitations \cite{BoschiFilho:2012xr}, this model seems to does not work well for higher spin glueball states for both QCD beta functions studied, namely, the beta function with an IR fixed point at finite coupling and the beta function with a linear asymptotic behaviour.

In this work, in particular, for beta function with an IR fixed point at finite coupling, using the 4th set of parameters $k$, $\lambda_0$ and $\lambda_\ast$, shown in the table \ref{t9}, were found values for the masses of the glueball states, both for higher even and odd spins, comparable with those found in the literature. This same beta function still provides Regge trajectories, for both pomeron and odderon, as shown in the equations (\ref{pad}) and (\ref{oad}), respectively,  in agreement with \cite{Landshoff:2001pp}, for the pomeron, and in \cite{LlanesEstrada:2005jf} for the odderon within the non-relativistic constituent model.

This leaves us with a question: why the softwall models with or without anomalous corrections do not work for high spin glueballs but the model with dynamical corrections does? To answer this question we plot the effective potentials of the anomalous dynamical softwall model for high spins discussed in this work in Figure \ref{plot9}. The effective potentials are given by: 
\begin{equation}\label{sch_ano_p}
V_{even \, J} =  k^2 z^2 + \frac{15}{4z^2}  + 2k  + \frac{1}{z^2} \left[  4 + J + \beta'(\lambda) - \frac{2}{\lambda} \beta(\lambda)\right] \left[ J + \beta'(\lambda) - \frac{2}{\lambda} \beta(\lambda)\right]\,,
\end{equation}
\begin{equation}\label{sch_ano_i}
V_{odd\, J} =  k^2 z^2 + \frac{15}{4z^2}  + 2k  + \frac{1}{z^2} \left[  6 + J + \beta'(\lambda) - \frac{2}{\lambda} \beta(\lambda)\right] \left[ 2 + J + \beta'(\lambda) - \frac{2}{\lambda} \beta(\lambda)\right]\,.
\end{equation}
 
One can see in Figure \ref{plot9} that the effectives potentials show an abrupt rise in the IR region (large $z$), simulating a wall or in this case a hardwall. As is well known, the hardwall model (with Neumann boundary condition) gives good results for even \cite{BoschiFilho:2005yh} and odd spins \cite{Capossoli:2013kb} related to the pomeron and to the odderon, respectively.  So, we conclude that the dynamical corrections lead to effective potentials that work like a hardwall at some finite value of the holographic coordinate $z$ implying good results for the glueball masses and Regge trajectories. 
Similar results have also been recently found for the (non-anomalous) dynamical softwall model \cite{Capossoli:2016kcr} and for a modified (analytical) softwall model \cite{Capossoli:2015ywa}.

\begin{acknowledgments}
H.B.-F. is partially supported by CNPq and E.F.C. by CNPq and FAPERJ, Brazilian agencies. D.L. is supported by the China Postdoctoral Science Foundation. 
\end{acknowledgments}


\begin{thebibliography}{99}

\bibitem{Landshoff:2001pp} 
  P.~V.~Landshoff,  ``Pomerons,''   hep-ph/0108156.
\bibitem{Cudell:2001ii} 
  J.~R.~Cudell and G.~Soyez,
  ``Does F(2) need a hard pomeron?,''
  Phys.\ Lett.\ B {\bf 516}, 77 (2001)  [hep-ph/0106307].
\bibitem{Levin:1998pk} 
  E.~Levin,  ``An Introduction to pomerons,''  hep-ph/9808486.
\bibitem{Meyer:2004jc} 
  H.~B.~Meyer and M.~J.~Teper,
  ``Glueball Regge trajectories and the pomeron: A Lattice study,''
  Phys.\ Lett.\ B {\bf 605}, 344 (2005)
  [hep-ph/0409183].
\bibitem{Chen:2005mg} 
  Y.~Chen {\it et al.},
  ``Glueball spectrum and matrix elements on anisotropic lattices,''
  Phys.\ Rev.\ D {\bf 73}, 014516 (2006)
  [hep-lat/0510074].
\bibitem{LlanesEstrada:2005jf} 
  F.~J.~Llanes-Estrada, P.~Bicudo and S.~R.~Cotanch,
  ``Oddballs and a low odderon intercept,''
  Phys.\ Rev.\ Lett.\  {\bf 96}, 081601 (2006)
  [hep-ph/0507205].
\bibitem{Maldacena:1997re} 
  J.~M.~Maldacena,
  ``The Large N limit of superconformal field theories and supergravity,''
  Adv.\ Theor.\ Math.\ Phys.\  {\bf 2}, 231 (1998)
  [hep-th/9711200].
\bibitem{Gubser:1998bc} 
  S.~S.~Gubser, I.~R.~Klebanov and A.~M.~Polyakov,
  ``Gauge theory correlators from noncritical string theory,''
  Phys.\ Lett.\ B {\bf 428}, 105 (1998)
  [hep-th/9802109].
\bibitem{Witten:1998qj} 
  E.~Witten,
  ``Anti-de Sitter space and holography,''
  Adv.\ Theor.\ Math.\ Phys.\  {\bf 2}, 253 (1998)
  [hep-th/9802150].
\bibitem{Witten:1998zw} 
  E.~Witten,
  ``Anti-de Sitter space, thermal phase transition, and confinement in gauge theories,''
  Adv.\ Theor.\ Math.\ Phys.\  {\bf 2}, 505 (1998)
  [hep-th/9803131].
\bibitem{Aharony:1999ti} 
  O.~Aharony, S.~S.~Gubser, J.~M.~Maldacena, H.~Ooguri and Y.~Oz,
  ``Large N field theories, string theory and gravity,''
  Phys.\ Rept.\  {\bf 323}, 183 (2000)
  [hep-th/9905111].
\bibitem{Polchinski:2001tt} 
  J.~Polchinski and M.~J.~Strassler,
  ``Hard scattering and gauge / string duality,''
  Phys.\ Rev.\ Lett.\  {\bf 88}, 031601 (2002)
  [hep-th/0109174].
\bibitem{Polchinski:2002jw} 
  J.~Polchinski and M.~J.~Strassler,
  ``Deep inelastic scattering and gauge / string duality,''
  JHEP {\bf 0305}, 012 (2003)
  [hep-th/0209211].
\bibitem{BoschiFilho:2002vd} 
  H.~Boschi-Filho and N.~R.~F.~Braga,
  ``Gauge / string duality and scalar glueball mass ratios,''
  JHEP {\bf 0305}, 009 (2003)
  [hep-th/0212207].
\bibitem{BoschiFilho:2002ta} 
  H.~Boschi-Filho and N.~R.~F.~Braga,
  ``QCD / string holographic mapping and glueball mass spectrum,''
  Eur.\ Phys.\ J.\ C {\bf 32}, 529 (2004)
  [hep-th/0209080].
\bibitem{Karch:2006pv} 
  A.~Karch, E.~Katz, D.~T.~Son and M.~A.~Stephanov,
  ``Linear confinement and AdS/QCD,''
  Phys.\ Rev.\ D {\bf 74}, 015005 (2006)
  [hep-ph/0602229].
\bibitem{Colangelo:2007pt}
  P.~Colangelo, F.~De Fazio, F.~Jugeau and S.~Nicotri,
  ``On the light glueball spectrum in a holographic description of QCD,''
  Phys.\ Lett.\ B {\bf 652} (2007) 73
  [hep-ph/0703316].
\bibitem{Forkel:2007tz} 
  H.~Forkel, M.~Beyer and T.~Frederico,
  ``Linear meson and baryon trajectories in AdS/QCD,''
  Int.\ J.\ Mod.\ Phys.\ E {\bf 16}, 2794 (2007)
  [arXiv:0705.4115 [hep-ph]].
\bibitem{Capossoli:2015ywa} 
  E.~F.~Capossoli and H.~Boschi-Filho,
  ``Glueball spectra and Regge trajectories from a modified holographic softwall model,''
  Phys.\ Lett.\ B {\bf 753}, 419 (2016) 
 [arXiv:1510.03372 [hep-ph]].
\bibitem{Gursoy:2007cb} 
  U.~Gursoy and E.~Kiritsis, ``Exploring improved holographic theories for QCD: Part I,''   JHEP {\bf 0802}, 032 (2008)   [arXiv:0707.1324 [hep-th]], U.~Gursoy, E.~Kiritsis and F.~Nitti, ``Exploring improved holographic theories for QCD: Part II,''   JHEP {\bf 0802}, 019 (2008)  [arXiv:0707.1349 [hep-th]], U.~Gursoy, E.~Kiritsis, L.~Mazzanti, G.~Michalogiorgakis and F.~Nitti, ``Improved Holographic QCD,''  Lect.\ Notes Phys.\  {\bf 828}, 79 (2011)  [arXiv:1006.5461 [hep-th]].
\bibitem{BoschiFilho:2012xr} 
  H.~Boschi-Filho, N.~R.~F.~Braga, F.~Jugeau and M.~A.~C.~Torres,
  ``Anomalous dimensions and scalar glueball spectroscopy in AdS/QCD,''
  Eur.\ Phys.\ J.\ C {\bf 73}, 2540 (2013)
  [arXiv:1208.2291 [hep-th]].

\bibitem{Narison:1988ts} 
  S.~Narison and G.~Veneziano,
  ``{QCD} Tests of $G$ (1.6) = Glueball,''
  Int.\ J.\ Mod.\ Phys.\ A {\bf 4}, 2751 (1989).
\bibitem{Gubser:2008yx} 
  S.~S.~Gubser, A.~Nellore, S.~S.~Pufu and F.~D.~Rocha,
 ``Thermodynamics and bulk viscosity of approximate black hole duals to finite temperature quantum chromodynamics,''
  Phys.\ Rev.\ Lett.\  {\bf 101}, 131601 (2008)
  [arXiv:0804.1950 [hep-th]].

\bibitem{deTeramond:2005su} 
  G.~F.~de Teramond and S.~J.~Brodsky,
  ``Hadronic spectrum of a holographic dual of QCD,''
  Phys.\ Rev.\ Lett.\  {\bf 94}, 201601 (2005)
  [hep-th/0501022].
\bibitem{BoschiFilho:2005yh} 
  H.~Boschi-Filho, N.~R.~F.~Braga and H.~L.~Carrion,
  ``Glueball Regge trajectories from gauge/string duality and the Pomeron,''
  Phys.\ Rev.\ D {\bf 73}, 047901 (2006)
  [hep-th/0507063].
\bibitem{Capossoli:2013kb} 
  E.~F.~Capossoli and H.~Boschi-Filho,
  ``Odd spin glueball masses and the Odderon Regge trajectories from the holographic hardwall model,''
  Phys.\ Rev.\ D {\bf 88}, no. 2, 026010 (2013)
  [arXiv:1301.4457 [hep-th]].
%
\bibitem{Zeng:2008sx} 
  D.~f.~Zeng,
  ``Heavy quark potentials in some renormalization group revised AdS/QCD models,''
  Phys.\ Rev.\ D {\bf 78}, 126006 (2008)
  [arXiv:0805.2733 [hep-th]].
\bibitem{Ryttov:2007cx} 
  T.~A.~Ryttov and F.~Sannino,
  ``Supersymmetry inspired QCD beta function,''
  Phys.\ Rev.\ D {\bf 78}, 065001 (2008)
  [arXiv:0711.3745 [hep-th]].
\bibitem{Alanen:2009na} 
  J.~Alanen and K.~Kajantie,
  ``Thermodynamics of a field theory with infrared fixed point from gauge/gravity duality,''
  Phys.\ Rev.\ D {\bf 81}, 046003 (2010)
  [arXiv:0912.4128 [hep-ph]].
\bibitem{Shock:2006gt} 
  J.~P.~Shock, F.~Wu, Y.~L.~Wu and Z.~F.~Xie,
  ``AdS/QCD Phenomenological Models from a Back-Reacted Geometry,''
  JHEP {\bf 0703}, 064 (2007)
  [hep-ph/0611227].
  
\bibitem{White:2007tu} 
  C.~D.~White,
  ``The Cornell potential from general geometries in AdS / QCD,''
  Phys.\ Lett.\ B {\bf 652}, 79 (2007)
  [hep-ph/0701157].
\bibitem{Li:2013oda} 
  D.~Li and M.~Huang,
  ``Dynamical holographic QCD model for glueball and light meson spectra,''
  JHEP {\bf 1311}, 088 (2013)
  [arXiv:1303.6929 [hep-ph]].
\bibitem{Capossoli:2016kcr} 
  E.~F.~Capossoli, D.~Li and H.~Boschi-Filho,
  ``Pomeron and Odderon Regge Trajectories from a Dynamical Holographic Model,''  arXiv:1601.05114 [hep-ph].


\end{thebibliography}
\end{document}